\documentclass[11pt]{article}
\usepackage{graphicx}
\usepackage{textcomp}
\usepackage{xcolor}
\usepackage{url}
\usepackage{tabularx}
\usepackage{hyperref}
\usepackage{fullpage}
\usepackage{comment}
\usepackage[disable]{todonotes}
\let\svtodo\todo\renewcommand\todo[1]{\svtodo[inline]{#1}}

\begin{document}

\title{The Case for ABI Interoperability in a Fault Tolerant MPI \\
}

\author{
Yao Xu\\
Khoury Coll\hbox{.} of Comp\hbox{.} Sciences \\
Northeastern University\\
Boston, MA, USA \\
xu.yao1@northeastern.edu
\and
Grace Nansamba\\
College of Engineering \\
Tennessee Tech\hbox{.} University\\
Cookville, TN, USA \\
gnansamba42@tntech.edu
\and
Anthony Skjellum\\
College of Engineering \\
Tennessee Tech\hbox{.} University\\
Cookville, TN, USA \\
askjellum@tntech.edu
\and
Gene Cooperman\\
Khoury Coll\hbox{.} of Comp\hbox{.} Sciences \\
Northeastern University\\
Boston, MA, USA \\
gene@ccs.neu.edu
}

\date{}

\maketitle

\begin{abstract}
There is  new momentum behind an interoperable ABI for MPI, which will be a major component of MPI-5.  This
capability brings true separation of concerns to a running MPI computation.
The linking and compilation of an MPI application becomes completely
independent of the choice of MPI library.  The MPI application is compiled
once, and runs everywhere.

This ABI allows users to independently choose: the compiler for the MPI application; the MPI runtime library; and, with this work,
the transparent checkpointing package.  Arbitrary combinations of the above are supported.
The result is a ``three-legged stool'', which supports performance, portability, and resilience for long-running computations.

An experimental proof-of-concept is presented, using the MANA checkpointing
package and the Mukautuva ABI library for MPI interoperability.
The result demonstrates that the combination of an ABI-compliant MPI and transparent checkpointing can bring extra flexibility in portability and dynamic resource management at runtime without compromising performance. For example, an MPI application can execute and checkpoint
under one MPI library, and later restart under another MPI library.
The work is not specific to the MANA package, since the approach using Mukautuva
can be adapted to other transparent checkpointing packages.
\end{abstract}

\vskip16pt
\section{Introduction}

This work introduces the paradigm of a three-legged stool for interoperability based on a common ABI.  That common ABI brings the benefits of independent compilation of: (i)~the MPI application; (ii)~the MPI library; and (iii)~the chosen checkpointing package.  Earlier work~\cite{hammond2023mpi} had already demonstrated the benefits of the first two components:  independent compilation of the MPI application and the MPI library.

This work focuses on the feasibility of adding ABI interoperability to an existing transparent checkpointing package.
Transparent checkpointing enables applications to be checkpointed and restarted without modifications to their source code
We achieve transparent checkpointing with MPI-agnostic interoperability by using the Mukautuva wrapper library.
Mukautuva wrappers enable an existing MPI library to be ABI-compliant and integrate seamlessly.

In particular, the focus on \emph{transparency} for checkpointing is essential
for HPC centers, due to their many applications.  While specific HPC applications sometimes use a application-specific checkpointing~\cite{nicolae2021veloc},
it would be impractical to modify each such MPI application in an HPC center:
This would require reimplementation, recompilation, and re-testing
for each MPI implementation.
For HPC center, en even worse obstacle is the case of vendor-provided closed-source applications.  These cannot be modified and recompiled in an application-specific approach.
In contrast, a transparent approach also simplifies system administration: a sysadmin can ``press a button" and checkpoint the unmodified application, even closed-source ones, regardless of its underlying MPI implementation. And not only does transparent checkpointing support varying MPI implementations, but it allows for seamless migration of applications across heterogeneous clusters with varying interconnects.

Hence, transparent checkpointing fills out the paradigm of a three-legged stool for
interoperability across diverse HPC environments. The paradigm
integrates the MPI compiler, the MPI library, and the checkpoint package.
\begin{enumerate}
\item When the MPI application is compiled (for example, with \texttt{mpicc}), then the \texttt{mpi.h} file for the chosen MPI compiler must support an ABI standard interoperable with other MPI implementations  (a feature to be standardized in
MPI-5~\cite{hammond2025mpi-abi-wg}, but is currently pre-standard).
\item The MPI library itself must also support this \texttt{mpi.h} API through a new library whose ABI is common to other MPI implementations.
\item The transparent checkpointing package must also support this \texttt{mpi.h} API through modification of its wrapper functions to support the new ABI, which is common to other MPI implementations.  The ABI support for transparent checkpointing is accomplished in this work by integrating with the Mukautuva library~\cite{hammond2024mukautuva} in a manner that can easily be adapted to other transparent checkpointing packages.
\end{enumerate}


\medskip\noindent\textbf{Motivation for ABI-compliant checkpointing:}
The motivation for including transparent checkpointing for ABI interoperability is precisely the same as for the original ABI proposal~\cite{hammond2023mpi}:  portability, compatibility and performance optimization.  That proposal highlights the benefits of a ``drop-in'' replacement, in which an existing MPI library is replaced with a new MPI library or new library version.  The new library supports linking using the same \texttt{mpi.h} file that was originally used.  The original MPI library must often be replaced when the hardware platform changes underneath, or when the software configuration changes (e.g., an incompatible RPATH variable).  Importantly, this change does administrative privilege, on account of the common ABI.

Thus, ABI-compliant support in a checkpointing package allows the user or sysadmin to migrate a running MPI computation to a new cluster, even though the new cluster may require a customized MPI library to support a new hardware platform or new hardware interconnect.  The reasons for migration may load balancing, system shutdown (e.g., electricity shutdown due to forest fires), or simply the desire to migrate a long-running computation to a new MPI library with an improved algorithm internal to the MPI library, or support for a newer hardware interconnect.

Still another motivation is improved MPI process layout across nodes of a cluster.  For example, assume a traditional MPI cluster, alongside a new, large-memory cluster with more cores. Then  processes with frequent communication can be migrated to a single node, where the local, ABI-compliant MPI library can enhance communication through shared memory.
(Note that transparent checkpointing packages ensure that there is no pending inter-process communication at checkpoint time~\cite{xu2024enabling}, and so a new ABI-compliant MPI library is free to use enhanced shared-memory communication for greater efficiency.)  Alternatively, a low-resource computation can be moved away from the high-memory cluster.

\medskip\noindent\textbf{ABI-compliant checkpointing as part of the three-legged stool:}
There are two major benefits for including transparent checkpointing in this
``three-legged stool''.
First, once a developer
has configured and compiled their MPI application, they can then make
an independent choice both of their preferred MPI implementation and
their preferred transparent checkpointing package.  In particular,
HPC sites can provide closed-source MPI applications from vendors,
while still independently changing to a new MPI library or a new transparent
checkpointing package.  Further, this aids efforts at delivering
complex MPI applications within containers.

A second benefit of the ``three-legged stool'' is that it enables easy migration of
an MPI computation to another cluster, perhaps while optionally
changing to a new MPI library.
The choice of new MPI library can be made for almost any reason.
These reasons include: migration to a less heavily loaded cluster, which
has a different preferred MPI library; trading off the performance of
one MPI implementation for better instrumentation and logging; or a
better mapping of the MPI processes to the hardware topology.

To demonstrate the potential of this paradigm, experiments are presented in which a transparent checkpoint is created with an MPICH implementation of MPI and restarted on Open~MPI, and vice versa, using the HPC-friendly CentOS~7 operating system and the Mukautuva library. This scenario illustrates how transparent checkpointing simplifies application portability and enables migration between clusters with similar operating systems but differing interconnects and different preferred MPI libraries. 

This work presents three points of novelty:
\begin{enumerate}
    \item A motivation was presented for introducing ABI interoperability into transparent checkpointing for MPI.
    \item A detailed design for ABI interoperability of transparent checkpointing was presented.
    \item An experimental evaluation confirms the practicality of ABI interoperability for transparent checkpointing, by using the MANA checkpointing package and the Mukautuva ABI interoperability library.
\end{enumerate}

 In the rest of this paper, Section~\ref{sec:related-work} presents the related work.
 Section~\ref{sec:background} further describes MANA, DMTCP and Mukautuva, and their relation to transparent checkpointing. Section~\ref{sec:three-legs} describes the existing state-of-the-art for an MPI ABI interoperability, including: the proposed MPI ABI standard; several implementations of wrapper/adapter libraries for MPI; and the ABI interface of the MANA package for checkpointing MPI.
Section~\ref{sec:experiments} provides an experimental evaluation of MANA and Mukautuva.
And Sections~\ref{sec:conclusion} and~\ref{sec:future-work} provide a conclusion and a summary of future work.

\section{Related Work}
\label{sec:related-work}

The Application Binary Interface (ABI) for MPI was presented as a solution to some issues of MPI such as the need for recompilation for new implementations, lack of standardization, and portability \cite{lindahl2005caseforABI}.  These issues make cross-implementation support harder, thus making the case for an MPI ABI by Lindahl~\cite{lindahl2005caseforABI}.
William Gropp~\cite{gropp2002building} proposed the use of generic wrappers and common object size to handle objects across different MPI implementations to create library components that can use any MPI. 

There are several systems/libraries that interpose between an application and an underlying MPI implementation or other underlying library, some of which are briefly discussed in this section.
In just the last three years, there have been at least five projects in this direction.  Mukautuva~\cite{hammond2024mukautuva} and MPItrampoline~\cite{Schnetter2022MPItrampoline} are two implementations embodying this approach.
Other efforts at dynamic translation through wrapper libraries include MPI\_Adapter~\cite{sumimoto2024mpi,sumimoto2009seamlessABI} for containers, Wi4MPI~\cite{Wi4MPI} for on-the-fly translation, and MPI Dialect~\cite{MLIRMPIDialect2023} for LLVM.
The first two efforts are interesting, in that instead of proposing a single ABI, they build a wrapper or adapter library from information in the header file \texttt{mpi.h}, on the fly.

The MPI\_Adapter~\cite{sumimoto2024mpi,sumimoto2009seamlessABI} is an automatic ABI translation library designed to extract the function prototypes and definition, insert into a conversion skeleton code for function mapping and accurate object handling. The MPI\_Adapter uses dynamic linking and translation techniques (i.e., dlsys, dlopen, \texttt{LD\_PRELOAD} for function interception), thus ensuring that MPI applications remain agnostic regardless of the underlying MPI implementation. 

The MPI Dialect~\cite{MLIRMPIDialect2023} in  Low Level Virtual Machine (LLVM) represents a modeling framework within LLVM  that aims to provide a standardized interface for interacting with MPI functionality at a lower level. It is intended to serve as an interfacing layer targeted by higher-level dialects within LLVM, abstracting away differences in the ABI, across multiple MPI implementations. 

MPItrampoline~\cite{Schnetter2022MPItrampoline} is an ABI for MPI that offers an MPI implementation through the ABI. MPItrampoline itself does not execute MPI functions, but it routes them to an actual implementation through the ABI. This setup enables the creation of portable applications compatible with any MPI implementation. (For example, it facilitates efficient development of external Julia packages using Yggdrasil --- a set of recipes for building packages for Julia  with ``BinaryBuilder.jl'' for nearly any HPC system.) An MPI wrapper library supports MPItrampoline across different MPI installations, the wrapper requires compilation for each specific MPI installation.

Of course, these multiple efforts of the last three years do not occur in a vacuum.
The MPI forum includes a working group specifically for an ``Application Binary Interface (ABI)''~\cite{hammond2023mpi,mpi2024abi}.


{

Other checkpointing libraries for MPI include 
VeloC~\cite{nicolae2021veloc}, FTI~\cite{bautista2011fti} and SCR~\cite{SCR2010Moody}. The goal among these is to support and enhance fault-tolerant MPI by creating a saved state (using checkpoints) where an application can recover from in case of failure. They can work with non-fault-tolerant MPI implementations and also support fault-tolerant and aware extensions to MPI (e.g., ULFM~\cite{ULFM2013Wesley}, Reinit~\cite{georgakoudis2020reinit}).}
Unfortunately, those approaches require separate modification of each individual application, followed by re-compilation.  Hence, they do not support the paradigm of a three-legged stool for interoperability:  the independent compilation of application, MPI library, and checkpointing package.

The next section discusses MANA (and the DMTCP checkpointing platform), along with Mukautuva, in more detail.

\section{Background: MANA, DMTCP, Mukautuva and Transparent Checkpointing}
\label{sec:background}

In keeping with the three-legged stool paradigm, fault tolerance is important for long-running MPI applications.  This is especially important when the computation time exceeds the time of a single resource allocation (typically 48~hours) at an HPC site.
For these long-running applications, transparent checkpointing provides a clean solution, without requiring modifications to the application itself.
DMTCP~\cite{ansel2009dmtcp} (Distributed MultiThreaded Checkpointing) is a widely used checkpointing package, and is an example of this transparent checkpointing approach.

MANA~\cite{garg2019mana,xu2023implementation} (MPI-Agnostic Network-Agnostic MPI) is supported as a plugin within the DMTCP platform~\cite{ansel2009dmtcp}).  MANA's previous use of virtual~ids~\cite[Section~4]{xu2023implementation} already formed an MPICH-specific basis for separate compilation of the MPI application, the MPI library, and the checkpointing package.  That work and Mukautuva are the foundation for the revised virtual~id design discussed in this paper.
Currently, MANA must be recompiled to adapt to the particular mpi.h, but this constraint will be relaxed in the future.

\medskip\noindent\textbf{ABI-compliant checkpointing and adapter libraries:}
The full benefits of the three-legged stool paradigm are realized only when all three ``legs of the stool'' support interoperability.
The glue behind all of these efforts is a \emph{wrapper library}, sometimes called an \emph{adapter library}, which interposes between the MPI application (or the checkpointing library) and the MPI library.

All of the above efforts can be thought of as a process of virtualization.  The concrete data objects are embedded in the MPI library (communicators, groups, request objects, MPI operations, etc.), and the MPI API allows an MPI application to retain references to these data objects.  It is this observation that makes it possible for MPI wrapper libraries or adapter libraries to virtualize the interface to the data objects.

\medskip\noindent\textbf{MANA, DMTCP and adapter libraries:}
By serendipity, transparent checkpointing faced exactly the same issue, and in 2009, DMTCP~\cite{ansel2009dmtcp} produced an analogous solution.  This was later generalized under the approach of \emph{process virtualization}~\cite{arya2016design}.  In this approach, an application saves a \emph{virtual} reference to the concrete data object in the Linux kernel.  MANA for MPI was later built on the DMTCP platform.  MANA adopted the process virtualization approach in a new framework called \emph{split processes}~\cite{garg2019mana}.  In this MANA framework, the MPI application embeds a virtual reference to the concrete data object in the MPI library.

The MANA framework for virtualization must not only support these virtual references to concrete MPI objects, but must also save and restore application memory.
MANA saves only the MPI application memory, and not the MPI data objects in the MPI library.
Just as DMTCP saves only the user application, and not the Linux kernel objects, MANA saves only the MPI application, and not the MPI data objects in the MPI library.  Later, on restart, the user application (for DMTCP or MANA) is restored to memory, its virtual reference to an underlying data object is used to recreate a semantically equivalent data object.

Until recently, MANA was only able to support a limited number of MPI implementations
due to the sheer variety of MPI implementations and due to limited developer resources.  In particular, when choosing virtual references, MANA hard-wired its implementation of virtual references to reflect the MPICH ABI,
including the choice of using a pointer or index into a table for resource references.
Later, MANA generalized its design using \emph{virtual ids}, so that at compile time, MANA would
be compiled with virtual references to MPI objects based either on the design of MPICH or Open MPI.
With this work, a strategic change has occurred: it is now possible for ABI-compliant implementations to work immediately without re-implementation, re-compilation, or re-testing.
To achieve this goal, MANA had simply to evolve to use the ABI instead of its own ABI-like infrastructure.

\medskip\noindent\textbf{Mukautuva:}
Finally, Mukautuva~\cite{hammond2024mukautuva} was developed in 2023 as a demonstration of the feasibility of a single ABI for the MPI compiler (in particular, for \texttt{mpi.h}) and as an ABI-compliant wrapper library to be integrated with an MPI library implementation.  Later, Hammond \hbox{et al.}~\cite{hammond2023mpi} reported on issues of concern for implementation of an ABI-compliant wrapper library, and demonstrated proof-of-concept experiments using Mukautuva.  Some examples of issues that were described are:  fields of the MPI status object; MPI datatype handles; certain Fortran constants for MPI, whose values may only be known at runtime; and packaging of MPI applications for different Linux distributions.



\vskip16pt
\section{The Three-legged Stool: Application ABI, MPI Library, and Checkpointing Package}
\label{sec:existing-proposals}
\label{sec:three-legs}

In this section, we discuss the proposed ABI standard of the MPI forum (Section~\ref{sec:application}),
the several implemented wrapper libraries for MPI libraries (Section~\ref{sec:library}),
and the implemented ABI for checkpointing packages (Section~\ref{sec:checkpointing}).
Currently, there is only one checkpointing package supporting an MPI ABI (MANA), but the same principles could be applied and implemented in any other transparent checkpointing package.

\bigskip
\subsection{The MPI Working Group ABI Proposal}
\label{sec:application}

The MPI ABI working group~\cite{hammond2023mpi} provides details about the ABI and explains the intended purpose of a standard ABI. At the time of this writing, those details are organized as github issues in a github repository.  The working group study is also informed by the design of concrete library implementations, such as Mukautuva.

The proposed components of the ABI include: MPI integer types, the status object, opaque handles, and values of MPI constants. 
Unlike the Application Programming Interface (API), which only defines how functions are called, the ABI defines how data is represented in memory.
By establishing a standard MPI ABI, it becomes possible for applications compiled with one MPI implementation to run with another, ensuring compatibility across different systems.
It is designed to support use cases such as direct MPI calls from third-party languages and interoperability between different MPI implementations.

\todo{Tony:  This paragraph is from the original EuroMPI paper.  Does it still fit? ---~Gene and Yao}
One challenge is in deciding how to handle deprecated functions within the standardized ABI. Two proposed options are to either exclude deprecated functions entirely or to remove them from the standardized ABI. While the former option allows implementations to retain deprecated functions for backward compatibility, it may introduce inconsistencies. Conversely, removing all deprecated functions ensures a streamlined ABI but could disrupt existing codebases reliant on these functions. 

\bigskip
\subsection{The MPI Library:  Multiple ABI Implementations}
\label{sec:library}

In this section, we discuss the existing ABI efforts, highlighting the major design features such as the split process, and the limitations associated with each of the ABIs. 

\medskip
\subsubsection{Mukautuva}\hfill\newline
Mukautuva is a compatibility layer for the different MPI implementations and prototypes for the ABI proposal being developed for the MPI Forum prototype designed by Hammond et \hbox{al.}~\cite{hammond2024mukautuva}. It includes two shared libraries: one provides MPI interface symbols, while the other delivers the core implementation, which is compiled against MPICH or Open MPI to facilitate the underlying functionality. At runtime, Mukautuva selects the appropriate implementation and activates it accordingly. 

\medskip
\subsubsection{Wi4MPI (Wrapper Interface for Multiple MPIs)}\hfill\newline
Wi4MPI~\cite{Wi4MPI} dynamically translates the ABI from the MPI library used during application compilation to a different MPI library available at runtime to overcome issues of ABI incompatibility across MPI libraries. The key features of Wi4MPI are the wrapper interface, on-the-fly dynamic translation, and runtime environment integration. This is done by either the preload version where it intercepts and translates MPI calls dynamically at runtime or the interface version where Wi4MPI acts like the MPI library itself. It uses a prefix system to differentiate between MPI calls from the application and from the runtime. Symbol overloading, hash tables, and thread safety using spinlocks are also important features.

Wi4MPI offers a streamlined solution for managing MPI library compatibility within containers for emerging workflows, ensuring flexibility, and performance optimization. The injection mechanism allows Wi4MPI to be dynamically integrated into containers at launch time. The injection mechanism enables translation using a function description file that includes the function signatures of the MPI interface, and the mapper description file that specifies how arguments are translated between the original library and the target library.
Some of the limitations of Wi4MPI include high overhead for small messages, partial support for MPI thread multiple, and limited processor family support.

\medskip
\subsubsection{Other ABI Translation Layers}\hfill\newline
For a discussion of other translation layers and adapter libraries, see the related work (Section~\ref{sec:related-work}).

\bigskip
\subsection{The Checkpointing ABI:  MANA}
\label{sec:checkpointing}
\label{sec:mana}

MANA is a transparent and efficient checkpointing tool that is agnostic to both MPI implementations and network configurations~\cite{garg2019mana,xu2023implementation}.
The design architecture for MANA is based on the \emph{split process}; split processes are a technique used to enable transparent checkpointing of MPI applications while avoiding the complexities of saving and restoring low-level hardware interactions.
This approach was recently extended to support both blocking and non-blocking MPI collective communication, using a topological sort approach with very low runtime overhead~\cite{xu2024enabling,xu2024arxivEnabling}.
The split process approach involves dividing a single process into two independent programs: the upper-half program, which is the MPI application itself, and a lower-half program, which includes the MPI library. In the upper half, the MPI application is compiled with regular mpicc. The libmana.so library of MANA is loaded using LD\_PRELOAD to intercept MPI function calls in the upper half. Wrapper functions in the libmana.so then invoke corresponding functions in the actual MPI library residing in the lower-half program. 

The MANA design, like Mukautuva~\cite{hammond2024mukautuva} and Wi4MPI~\cite{Wi4MPI}, employs a wrapper-based interface.  The wrapper design is forced on MANA due to its split process architecture, and the need for the upper-half MPI application to call the lower-half MPI library.  The initial MANA implementation~\cite{garg2019mana} was designed around the MPICH family's choice to implement the various MPI data structures.  A later version of the MANA project added \emph{virtual ids} for MPI objects in 2023, thus allowing MANA to support other MPI implementations.  This was reported on in~\cite{xu2023implementation}, where MANA support for HPE/Cray MPICH, Open MPI, and ExaMPI were demonstrated. In this work, we report on a revised version of MANA, which eliminates the original requirement~\cite{garg2019mana,xu2023implementation} for a statically linked MPI library in the lower half.
The revised version can currently be found at~\cite{mana2024dynamic}, and will soon replace the current main branch of MANA~\cite{mana2024main}.

While virtual ids enabled MANA support for multiple MPI implementations, it was still required to recompile MANA for each MPI implementation, based on the \texttt{mpi.h} API and internal data structures.  The internal data structures mattered because the MPI types exposed by \texttt{mpi.h} were variously implemented as pointers, indexes in a table, or other opaque data structures.

\begin{figure}[h!t]
\centering
\includegraphics[width=0.5\columnwidth]{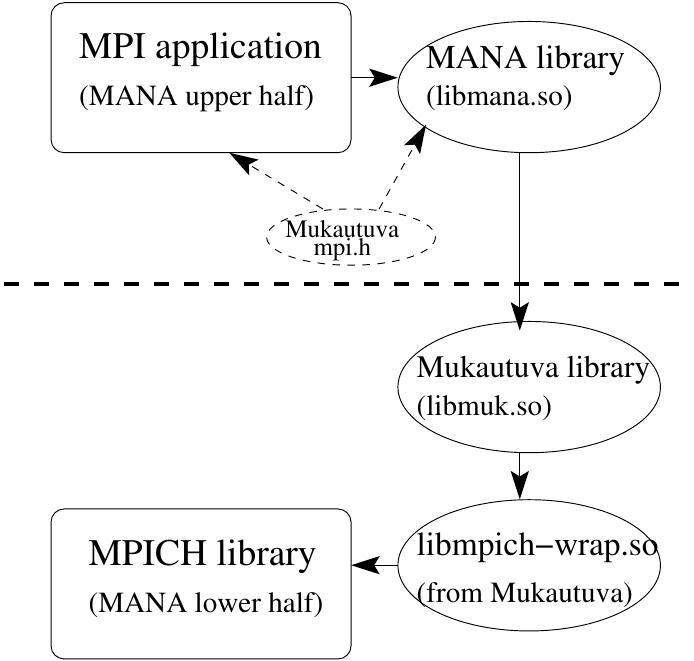}
\caption{In this example, MANA is running over MPICH, although other MPI libraries are possible.  The \texttt{libmuk.so} library dynamically detects the MPICH library at runtime, and loads \texttt{libmpic-wrap.so}.  The horizontal dotted line separates the upper and lower half in the split process approach.  At checkpoint time, only the memory of the upper half is saved.  At restart time, (i)~the memory of the upper half is restored; (ii)~a fresh copy of the lower half (including Mukatauva) is launched; and the wrapper functions of \texttt{libmana.so} are bound to the wrapper functions of \texttt{libmuk.so}.}
\label{fig:mana-mukautuva}
\end{figure}

By linking MANA to a single Mukautuva wrapper library (\texttt{libmuk.so}), along with Mukautuva's own \texttt{mpi.h} API, this work demonstrate support for a single ABI for MPI, that of Mukautuva.
See Figure~\ref{fig:mana-mukautuva} for an overview of MANA's integration with Mukautuva.
The Mukautuva wrapper library does the ``heavy lifting'', by translating the MPI types and MPI constants from the Mukautuva interface to the particular MPI library interface.  At runtime, the Mukautuva library, libmuk.so, dynamically loads \texttt{libmpich-wrap.so} (also provided by Mukautuva).  MANA need only support a single Mukautuva interface.  In a future MPI ABI standard, each MPI implementation will absorb \texttt{libmpich-wrap.so} or the equivalent into its own implementation.

As a result, MANA is compiled just once, and is then re-used to support MPICH, Open MPI, or some other MPI implementation that supports the Mukautuva interface.

Section~\ref{sec:experiments} shows the runtime performance: for native HPE/Cray MPI and Open MPI; for MANA's support of HPE/Cray MPI and Open MPI with Virtual ID; and for MANA's support of both HPE/Cray MPI and Open MPI (using Mukatuva).

\vskip16pt
\section{Experimental evaluation of MANA using the Mukautuva interoperable ABI}
\label{sec:experiments}
Experiments with MANA were performed on the Discovery cluster of Northeastern University at Massachusetts Green High Performance Computing Center (MGHPCC), using four compute nodes with 48 Intel Xeon E5-2690 v3 CPUs. Compute nodes are connected by 10 GbE interconnect. Both MPICH~3.3.2 and Open MPI~3.1.2 were used to show MANA's interoperability. \textbf{mpicc} for both MPICH and Open~MPI
is based on gcc-11.1.0. The Linux operating system is CentOS~7, with Linux kernel~3.10.0.
All experiments are repeated 5~times.

\subsection{Runtime overhead of MANA and Mukautuva}
\label{sec:microbenchmark}
In this subsection, we use OSU Micro-benchmark~7.5~\cite{osu-microbenchmark} to show that the addition of MANA and Mukautuva does not bring extra runtime overhead. We picked three common MPI functions with different communication patterns: \texttt{MPI\_Alltoall}, \texttt{MPI\_Bcast}, and \texttt{MPI\_Allreduce}.

\begin{figure}[h!t]
\centering
\includegraphics[width=0.5\columnwidth]{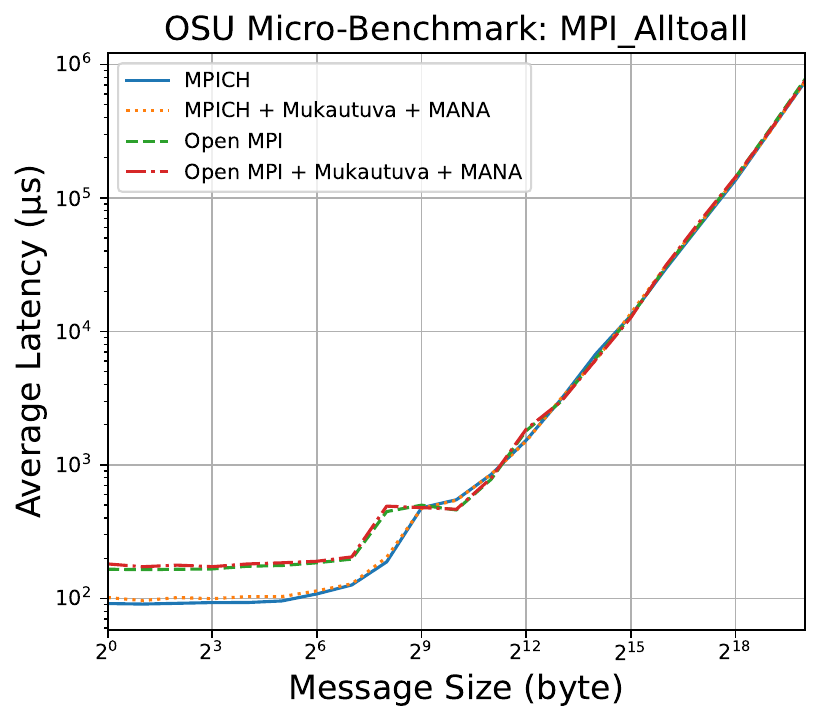}
\caption{Median Latency of MANA for OSU Micro-benchmarks for \texttt{MPI\_Alltoall}. Results are in two groups: Cray MPICH and Open MPI. In each group, there are different configurations: native, and with both Mukautuva and MANA. Note that times for each group are almost identical, indicating that MANA and Mukautuva add little runtime overhead. }
    
\label{fig:alltoall}
\end{figure}

\begin{figure}[h!t]
\centering
\includegraphics[width=0.5\columnwidth]{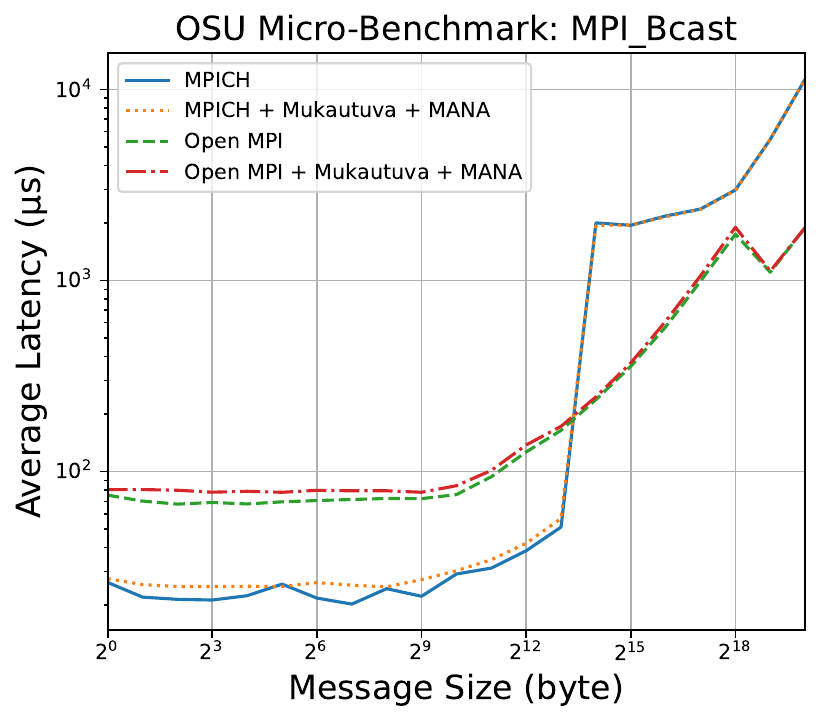}
\caption{Median Latency of MANA for OSU Micro-benchmarks for \texttt{MPI\_Bcast}. Results are in two groups: Cray MPICH and Open MPI. In each group, there are different configurations: native, and with both Mukautuva and MANA. \texttt{MPI\_Bcast} is more efficient than \texttt{MPI\_Alltoall}. Therefore, we can observe some small runtime overhead with small message sizes. This runtime overhead is due to the lack of a kernel feature in Discovery's old Linux kernel that is required by MANA to improve runtime efficiency.} 
\label{fig:bcast}
\end{figure}

\begin{figure}[h!t]
\centering
\includegraphics[width=0.5\columnwidth]{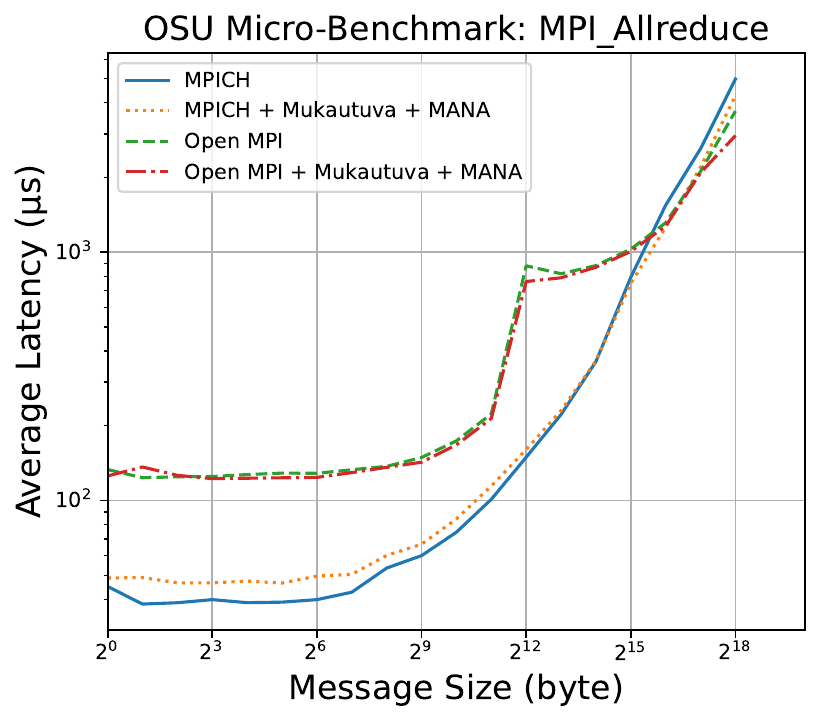}
\caption{Median Latency of MANA for OSU Micro-benchmarks for \texttt{MPI\_Allreduce}. Results are in two groups: Cray MPICH and Open MPI. In each group, there are different configurations: native, and with both Mukautuva and MANA. We can observe the same runtime overhead with small message sizes. In addition, the result of Mukautuva and MANA outperformed the native result in some cases because of the larger standard deviation in testing results} 
\label{fig:allreduce}
\end{figure}

Figures~\ref{fig:alltoall}$\,$, \ref{fig:bcast}, and \ref{fig:allreduce} show the absolute time (latency) of each MPI communication call on four compute nodes with 48 MPI processes. The latencies and message sizes are in the log-log scale.
The runtime overhead of Mukautuva and MANA is represented as the gap between lines of the same MPI library, MPICH or Open MPI.

Among all three MPI functions, \texttt{MPI\_Alltoall} (Figure~\ref{fig:alltoall}) is the most network-intensive one. From Figure~\ref{fig:alltoall}, we can see that although Open MPI and MPICH have different latencies according to the message sizes, there's barely any gap between the native performance and Mukautuva and MANA. The largest runtime overhead appears when the message size is 1 byte. In this case, the maximum runtime overhead is 10.9\%. Then the runtime overhead rapidly drops under 1\% as the message size grows.

In comparison, \texttt{MPI\_Bcast} (Figure~\ref{fig:bcast}) and \texttt{MPI\_Allreduce} (Figure~\ref{fig:allreduce}) are more efficient because there are fewer messages that need to be delivered. As a result, the runtime overhead of Mukautuva and MANA is more noticeable in Figure~\ref{fig:bcast} and Figure~\ref{fig:allreduce}, especially with smaller messages. The maximum runtime overhead is 17.2\%.

A major cause of Mukautuva and MANA's runtime overhead is the lack of a Linux kernel feature on Discovery: setting the FSGSBASE register directly in userspace. This feature was introduced in Linux kernel~5.9. MANA relies on setting the FSGSBASE register to context switch between the upper and lower half. Without this new kernel feature, MANA needs to use syscall to set the FSGSBASE register, which is expensive.

The overhead due to FSGSBASE is an artifact of the split process architecture used by MANA (see Figure~\ref{fig:mana-mukautuva}).  Even though the diagram represents a single process, the MPI application was loaded as part of a MANA upper half program, while the MPICH library was loaded as part of a MANA lower half program.  In Linux, the x86 CPU fs register is used as part of a pointer to the current thread of a process.  Because we have two programs that were independently loaded, the two programs are compiled separately, any function call from the upper half to the lower half must also change the thread context (\hbox{i.e.} the fs register).  Before Linux kernel~5.9, a user program could modify the fs register only by calling a kernel function, and the kernel call presents high overhead when calls to the lower half of the split process are frequent. 

\begin{figure*}[h!t]
\centering
\includegraphics[width=0.8\textwidth]{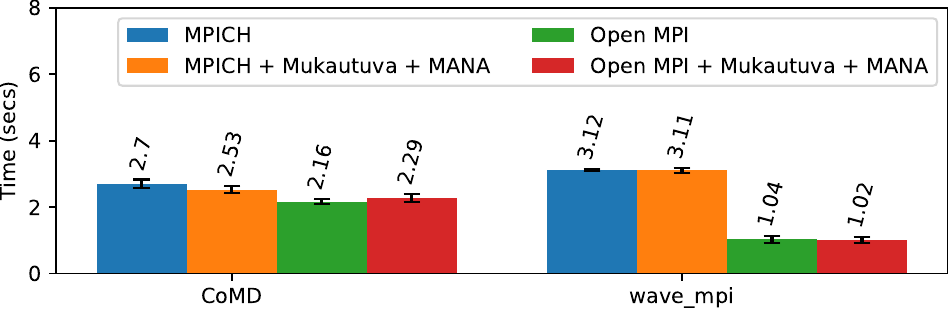}
\caption{Runtime performance of real-world MPI applications. Here we show the median of the completion time of both applications.  The error bars show the standard deviation.} 
\label{fig:real_world_app}
\end{figure*}

Note also that micro-benchmarks represent an absolute worst case for latency or runtime overhead.  The micro-benchmarks call the same MPI communication function repeatedly without any computational work between communications.  However, in a real-world program, there will be significant computation before calling an MPI communication function.  Therefore, the next section, ``Real-world Applications'' provides a more realistic estimate of the overhead that users of MANA+Mukautuva will see.

\subsection{Runtime Overhead of Real-world Applications}
In this subsection, we will showcase real-world MPI applications with Mukautuva and MANA. We chose CoMD~\cite{papa2001constrained} and wave\_mpi~\cite{burkardt2013wave} as our examples. We run both applications on four compute nodes with 48 MPI processes. Each application is compiled with MPICH, Open MPI, and Mukautuva.

Real-world MPI applications usually communicate less intensively compared to micro-benchmarks. Therefore, the runtime overhead is smaller than micro-benchmarks.

Figure~\ref{fig:real_world_app} shows the performance of two MPI applications. Compared to the micro-benchmarks, the new results exhibit larger standard deviations. From the graph, CoMD appears to run faster when using Mukautuva and MANA with MPICH. This is due to the unstable results. In most experiments we did, Mukautuva and MANA have the same performance as CoMD with native MPICH.

Similarly, when running CoMD with Open MPI, Mukautuva and MANA have 5\% runtime overhead. However, the standard deviation of Open MPI is 0.07~seconds, and the deviation of Mukautuva and MANA is 0.1~seconds.

For wave\_mpi, there's almost zero runtime overhead, both in comparing MPICH to MPICH+Mukautuva+MANA and in comparing Open~MPI to OPEN~MPI+Mukautuva+MANA.

\subsection{Launch with one MPI Implementation, Restart with Another one}
In this subsection, we will demonstrate checkpointing an MPI program using Open MPI, and later restarting with MPICH.

We modified the OSU Micro-benchmark for \texttt{MPI\_Alltoall} for this experiment. Before starting the measurement, the micro-benchmark program has a ``warm-up'' phase that performs some dummy communications. We modified the program so that it will sleep for 10~seconds after the warm-up phase. After launching the micro-benchmark program with Open MPI, we create a checkpoint using this time window.

\begin{figure}[h!t]
\centering
\includegraphics[width=0.5\columnwidth]{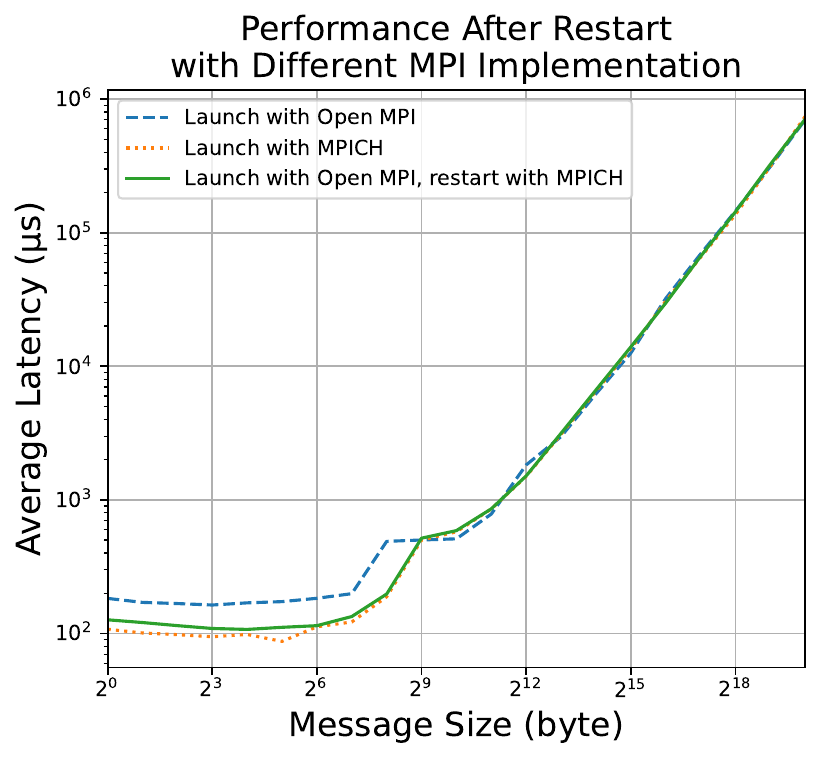}
\caption{Median Latency of MANA for OSU Micro-benchmarks for \texttt{MPI\_Alltoall}. The dashed blue line and dotted orange line represent the performance of Open MPI and MPICH with Mukautuva and MANA. The solid green line represents the performance after restarted with a different MPI implementation (MPICH) other than the implementation used for launch (Open MPI).} 
\label{fig:restart}
\end{figure}

After the checkpoint completes, the micro-benchmark program will resume running, and the latency results will be recorded (blue dashed line in Figure~\ref{fig:restart}). Later, we restart the saved checkpoint images with MPICH and collect the performance results (green solid line in Figure~\ref{fig:restart}). This process is seamless with the help of Mukautuva.

Finally, we launch the micro-benchmark program again with MPICH for comparison (orange dotted line in Figure~\ref{fig:restart}).

Figure~\ref{fig:restart} shows that the performance after restart is not affected even though the underlying MPI library has been switched from Open MPI to MPICH. There is some runtime overhead with small message sizes (up to 17\% overhead), but it shows almost the same pattern as results in subsection~\ref{sec:microbenchmark}.

\vskip16pt
\section{Conclusion}
\label{sec:conclusion}

This work presents the case for incorporating Application Binary Interface (ABI) support in transparent checkpointing mechanisms such as MANA.  This capability also significantly enhances both portability and interoperability.  The result is the ability to employ dynamic resource management at runtime, without compromising performance.

Experiments were presented employing the MANA package for transparent checkpointing, the Mukautuva adapter library, and two MPI implementations:  MPICH and Open~MPI.  Nevertheless, this proof-of-concept does not depend on any particular package.  For example,MPICH and Open~MPI are made ABI-compliant by using Mukautuva as a wrapper library (wrapping MPI functions).  Thus, the methodology is not specific to a particular MPI implementation.  Similarly, MANA was easily made ABI-compliant by using the Mukautuva wrapper library.  Thus, the methodology would be easily adapted to any transparent checkpointing package.

The adoption of an ABI in the MPI standard will soon facilitate seamless execution across different MPI implementations without requiring application recompilation.  This removes  
the dependency on specific MPI implementations, which would otherwise limit the portability and rehosting of HPC applications that are distributed in binary form.  This capability also eases the integration with productivity environments like Python and Rust, with the many pre-compiled parallel libraries in their respective ecosystems.

This work also highlights the functionality of current ABI implementations such as Mukautuva~\cite{hammond2024mukautuva}, MANA~\cite{xu2023implementation}, and Wi4MPI~\cite{Wi4MPI}, all of which aim to simplify the complexities associated with application portability across implementations on common platforms (e.g., x86-64 with Linux). 
By aligning transparent checkpointing with these proposed ABI advancements for MPI, this paper identifies how MPI applications can achieve new levels of flexibility. 
Additionally, the ongoing development of ABIs could usher in a new era of MPI usage in which developers can focus more on innovation and less on the intricacies of specific MPI implementations.
Our strategy calls for a harmonized approach where future ABI standard(s) not only ease the rehosting across different MPI implementations, but also integrate seamlessly with advanced fault tolerance mechanisms (e.g., transparent checkpointing).

\todo{WHAT IS THIS? [I think this is now stale text, since MANA/Mukautuva exists concretely today, and we have a new story.  But I'm keeping, above: ``Our strategy calls for a harmonized approach where future ABI standard(s) not only $\ldots$ advanced fault tolerance mechanisms (e.g., transparent checkpointing).'' ---~Gene] OLD TEXT: The paper elaborates on the future roadmap for ABIs in MPI, suggesting enhancements that further streamline the checkpointing process and eliminate the need for undesired application recompilations.}


Our results also demonstrate that MANA maintains low overhead while providing ABI interoperability across various MPI implementations. This confirms MANA's capability for achieving scalable ABI compatibility with minimal performance impact.  ABI standardization advances checkpointing in modern MPI environments, including productivity environments built on Python, Rust, etc.  

\vskip16pt
\section{Future work}
\label{sec:future-work}

 MANA has already been shown to work well with the Mukautuva ABI library.
 For simplicity and compatibility, MANA will eventually just use the MPI standardized ABI (as currently represented in Mukautuva and its updates) and become fully interoperable without application recompilation, for the sake of use cases involving rehosting, malleability, and/or fault tolerance.  

In the future, we will also explore the use of containerization techniques, such as Docker \cite{Docker} and Singularity \cite{kurtzer2017singularity}, to deploy MANA with ABI-compatible MPI libraries in containerized environments. We will evaluate the performance, scalability, and resource utilization of MANA within containerized deployments, considering factors such as overhead and isolation.

Note that MANA achieves its purposes solely by interposing on MPI library calls (see Figure~\ref{fig:mana-mukautuva}).  While MANA has been used primarily, to date, to support applications in C, C++ and Fortran, there is nothing intrinsically problematic in extending this support to interpreted ``productivity'' languages.  The underlying MANA platform is DMTCP, which already supports checkpointing of applications purely in Python and Julia.
We will conduct experiments to evaluate the interoperability of MANA with applications written in different programming languages, such as Python and Julia, by leveraging ABIs for direct MPI calls. We will measure the effectiveness of MANA in providing transparent checkpointing for applications developed using diverse programming paradigms such as those built on Python (including compiled variants) and Julia.

We plan to revisit the potential for Fortran support with MANA once the standardized ABI for MPI fully addresses Fortran.
 

Finally, understanding the detailed shape of the performance curves shown in Figures~\ref{fig:alltoall} and~\ref{fig:bcast} requires deeper instrumentation and code review of the underlying mechanisms leveraged by MANA.

\bibliographystyle{ieeetr}
\bibliography{references}

\end{document}